\documentclass{anabs}
\usepackage{graphicx}
\usepackage{times}
\Volume{0}              
\Year{2004}              
\Month{00}                
\Pagespan{000}{000}      

\begin{document}
\lhead[\thepage]{A.N. Author: Title}
\rhead[Astron. Nachr./AN~{\bf 325}, No. \volume\ (\yearofpublic)]{\thepage}
\headnote{Astron. Nachr./AN {\bf 325}, No. \volume, \pages (\yearofpublic) / {\bf DOI} 10.1002/asna.\yearofpublic1XXXX}

\title{HST/FOS Eclipse mapping of IP~Pegasi in outburst}

\author{R.K. Saito\inst{1}, R. Baptista\inst{1} and K. Horne\inst{2}}
\institute{
Departamento de F\'{\i}sica, Universidade Federal de Santa Catarina, Trindade, 
88040-900, Florian\'opolis, SC, Brazil
\and 
School of Physics and Astronomy, University of St. Andrews, KY16 9SS, Scotland, UK}

\date{Received (date); accepted (date); published online (date)}

\correspondence{saito@astro.ufsc.br}

\maketitle

\section{Introduction}

IP~Pegasi is a deeply eclipsing dwarf nova ($P_{orb}= 3.8$~hr) which 
shows 1-2 weeks-long, $\simeq 2$~mag outbursts every 60-120 days.
During outbursts, tidally induced spiral shocks form in its accretion 
disc as the disc expands and its outer parts feel more effectively the 
gravitational attraction of the companion star (Steeghs, Harlaftis \& Horne 1997).
Here we present the first results of a time-resolved ultraviolet (UV)
spectral mapping experiment of IP~Pegasi 9-13 days after onset of the 
May 1993 outburst.

\section{Data analysis and results}

Time-resolved eclipse spectroscopy covering 3 eclipses of IP~Peg was 
secured with the Faint Object Spectrograph onboard the Hubble Space 
Telescope on 1993 May 27-30 (G160L grating, $\lambda\ 
1100-2500$~\AA, $\Delta\lambda= 1.7$ \AA\ pixel$^{-1}$). 
The runs cover the eclipse cycles 22249 (IP5), 22252 (IP6) and 22263 
(IP7), according to the ephemeris of Wolf et al. (1993).

The out-of-eclipse UV spectra show prominent Ly$\alpha$,
Si\,IV $\lambda 1400$, C\,IV $\lambda 1550$ and He\,II $\lambda
1640$ emission lines as well as broad absorption bands possibly due to 
Fe\,II.
The spectra were divided into 27 narrow passbands and light curves were 
extracted for each one.
Maximum-entropy eclipse mapping techniques were used to solve 
for a map of the disc brightness distribution and for the flux of an
additional uneclipsed component in each band (Baptista \& Steiner 1993).
The line maps are brighter at disc centre, whereas the
continuum maps show a flat brightness distribution with no
pronounced central source.
The IP5 continuum maps show an asymmetric structure at a similar 
azimuth and radius as one of the spiral arms seen in Baptista, 
Harlaftis \& Steeghs (2000). 
The other spiral arm has possibly been lost because of the incomplete 
phase coverage of the light curves of this data set. 
The time sequence of eclipse maps show the progressive fading of 
both line and continuum emission as the outburst ends. 

Spatially resolved disc spectra show strong Si\,IV, C\,IV and
He\,II lines, which appear in emission everywhere in the disc.
The spectrum of the bright spot is different from the 
disc spectrum only in the outer disc regions, with no
evidence of a gas stream overflow. 
The lines are stronger in the inner disc regions and decrease in 
strength with increasing disc radius.
The FWHM of the C\,IV line is approximately constant with 
radius, in contrast to the expected $v \propto R^{-1/2}$ law for gas
in Keplerian orbits. The radial run of the FWHM of He\,II and Si\,IV
is also flatter than the Keplerian expectation.
This suggests that these lines do not arise in the disc photosphere,
but possibly in a disc chromosphere + wind.

Blackbodies and LTE H\,I emission spectra give poor fits to the slope of
the flat UV continuum in the spatially-resolved spectra and lead to 
inconsistent and uncomfortably high temperatures. 
We fitted Kurucz's (1979) stellar atmosphere models to the 
continuum emission at each disc radius in order to estimate the radial 
run of the T$_{\rm eff}$. For IP5 and IP6,
the temperatures range from 40000~K in the inner disc to 9000~K in
the outer disc regions and are reasonably well described by steady-state
disc models, respectively, of \.{M}$= 10^{-8.5}$ and $10^{-8}
\;{\rm M_{\odot} yr}^{-1}$. This is in marked contrast with the previous
optical results (Baptista, Haswell \& Thomas 2002), which points to flat 
brightness temperature distributions with temperatures in the range 
$5000-9000$~K.
The derived temperatures at the end of the outburst (IP7) are everywhere 
higher than the critical temperature above which the gas should remain
in a steady, high-mass accretion regime (Warner 1995).

\end{document}